\documentclass[aps,prd,preprint,groupedaddress,showpacs,eqsecnum,nofootinbib,floatfix]{revtex4}
\usepackage{graphicx,epsf,subfigure}
\graphicspath{{./plots/},{./diagrams/}}

 \newif\ifdraft
\drafttrue
\newif\ifpreprint
\preprinttrue

\def\fig#1{fig.~{\ref{#1}}}
\def\Fig#1{Fig.~{\ref{#1}}}
\def\figs#1#2{figs.~{\ref{#1}} and {\ref{#2}}}

\newcommand{\be}{\begin{equation}}
\newcommand{\ee}{\end{equation}}
\newcommand{\ba}{\begin{eqnarray}}
\newcommand{\ea}{\end{eqnarray}}

\newcommand{\BlackHat}{{\sc BlackHat}}
\newcommand{\SHERPA}{{\sc SHERPA}}
\newcommand{\AMEGIC}{{\sc AMEGIC++}}
\newcommand{\COMIX}{{\sc COMIX}}
\newcommand{\jetphox}{{\sc JetPhoX}}
\newcommand{\HTj}{H_T^{\rm jet}}

\def\Zjj{$Z\,\!+\,2$}

\def\Zjn{$Z\,\!+\,n$}
\def\Wjn{$W\,\!+\,n$}

\def\gjj{$\gamma\,\!+\,2$}
\def\gp{$\gamma\,\!+\,$}

\def\nub{\bar\nu}
\def\qb{\bar q}

\begin{document}
\hfuzz 10 pt


\ifpreprint
\noindent
UCLA/11/TEP/106 \hskip2.3cm
SLAC--PUB--14472
\hfill CERN--PH--TH/2011-125\\
IPhT--T11/146\hskip3.4cm
IPPP/11/29
\hfill NSF--KITP--11--093\\
SB/F/387-11
\fi

\title{Driving Missing Data at Next-to-Leading Order}

\author{Z.~Bern${}^a$, G.~Diana${}^b$,
L.~J.~Dixon${}^{c,d}$, F.~Febres Cordero${}^e$, 
S.~H{\"o}che${}^d$,  H. Ita${}^a$, D.~A.~Kosower${}^b$,
D.~Ma\^{\i}tre${}^{c,f}$ and K.~J.~Ozeren${}^a$}

\affiliation{
\centerline{${}^a$Department of Physics and Astronomy, UCLA,
Los Angeles, CA 90095-1547, USA} \\ 
\centerline{${}^b$Institut de Physique Th\'eorique, CEA--Saclay,
F--91191 Gif-sur-Yvette cedex, France} \\
\centerline{${}^c$Theory Division, Physics Department, CERN,
CH--1211 Geneva 23, Switzerland} \\
\centerline{${}^d$SLAC National Accelerator Laboratory,
Stanford University, Stanford, CA 94309, USA} \\
\centerline{${}^e$Departamento de F\'{\i}sica, Universidad
Sim\'on Bol\'{\i}var, Caracas 1080A, Venezuela} \\
\centerline{${}^f$Department of Physics, University of Durham,
Durham DH1 3LE, UK} }


\begin{abstract}
The prediction of backgrounds to new physics signals in topologies
with large missing transverse energy 
and jets is important to new physics searches at the LHC.
Following a CMS study, we investigate theoretical issues in using
measurements of \gjj-jet production to predict the irreducible
background to searches for missing energy plus two jets that
originates from \Zjj-jet production where the $Z$ boson decays to
neutrinos.  We compute ratios of \gjj-jet to \Zjj-jet production cross
sections and kinematic distributions at next-to-leading order in
$\alpha_s$, as well as using a parton shower matched to leading-order
matrix elements.  We find that the ratios obtained in the two
approximations are quite similar, making \gjj-jet production a
theoretically reliable estimator for the missing energy plus two jets
background.  We employ a Frixione-style photon isolation, but we also
show that for isolated prompt photon production at high transverse
momentum the difference between this criterion and the standard cone
isolation used by CMS is small.

\end{abstract}

\pacs{12.38.Bx, 13.85.Qk, 13.87.Ce\hspace{1cm}}

\maketitle


\renewcommand{\thefootnote}{\arabic{footnote}}
\setcounter{footnote}{0}

\section{Introduction}
\label{IntroSection}

The LHC era is now upon us, and the hunt for the mechanism of
electroweak symmetry breaking and new physics beyond the standard
model is underway. Typical signatures for supersymmetry and many other
new physics models include topologies with large missing transverse
energy (MET) accompanying jets (METJ).  The same signatures can easily
be mimicked by Standard Model processes, such as the production
of an electroweak boson decaying into neutrinos, in association with jets.
For the discovery of new physics in the early running of the LHC,
with only a few inverse femtobarns of integrated luminosity, it is
important to understand the Standard Model backgrounds to METJ searches.

Events containing a $Z$ boson and jets, with the $Z$ decaying into a
neutrino pair (METZJ), constitute an irreducible background to the
METJ signal. One can envisage using complementary approaches to
understanding this and other backgrounds: a direct theoretical prediction;
or data-driven approaches, which estimate the rate from measurements of other
processes (or possibly from other kinematic regions in the same process).
Data-driven techniques offer a powerful means of avoiding
theoretical uncertainties in background predictions, as well as cancelling
experimental systematics common to different processes.
However, such methods can require theoretical assistance, in order
to estimate the ``translation'' parameters from one process to another,
and their inherent uncertainties.  The question theorists can address
is how stable is the ratio of two processes to various theoretical
approximations.

In the case of the process $Z(\to\nu\nub)\,\!+\,$jets, the most
obvious choice of other process would be
$Z(\to\ell^+\ell^-)\,\!+\,$jets, {\it i.e.}~the production of a $Z$
boson in association with jets, where the $Z$ decays into a charged
lepton pair.  The production kinematics and dynamics of these two
processes are identical, so no theoretical input about QCD is
required, only knowledge of the $Z$ boson branching ratios.
Leptonically decaying $Z$ bosons have the drawback, however, of
offering very low event rates, less than a sixth of METZJ (per lepton
channel), even before imposing lepton rapidity cuts.  The paucity of
statistics has led experimenters to examine using other processes to
estimate METZJ rates and distributions.  The CMS collaboration has
studied~\cite{CMS-photon-note,CMSMET} the use of $W$ or photon
production in association with jets for estimating the METZJ
background.  The production of a $W$ in association with jets offers
an order of magnitude higher statistics than the leptonic $Z$ process;
the production of a prompt photon in association with jets,
sixteenfold higher statistics than leptonic $Z$ decays.  In addition,
$W$ production suffers from contamination from $t\overline t$ events,
and could well be affected by the same kinds of new physics the
experiments are seeking in the METZJ sample.  Reducing $t\overline t$
contamination requires selection cuts that enhance the photon
channel's advantage.

Use of either the $W$ or $\gamma$ plus jets processes requires
knowledge of their short-distance strong-interaction dynamics, as they probe
different combinations of the parton distributions and somewhat
different scales than $Z$ plus jets.
In the photon case, its masslessness also affects
distributions in important ways and requires theoretical input for any
comparison with massive boson production.  The study of both processes
is valuable, of course; the use of different processes allows for
cross checks and also presumably different sensitivity to whatever new
physics may be lurking in the data.

\begin{figure}[t]
\begin{center}
\includegraphics[clip,scale=0.45]{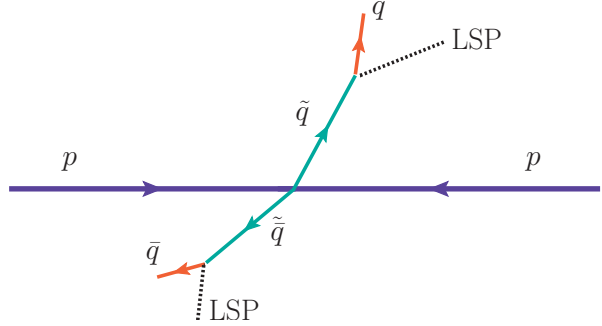}
\end{center}
\caption{Squark pair production illustrates a new physics process with
the signature of two jets plus MET.  Here each squarks decays to a quark
and the lightest neutralino; the escaping neutralinos generate the 
missing transverse energy.}
\label{fig:NewPhysicsExample}
\end{figure}

In this paper, we study the \gjj-jet and
$Z(\to\nu\nub)\,\!+\,2$-jet production processes.
The latter process is a background to new physics, such as the squark
pair-production process illustrated in \fig{fig:NewPhysicsExample}.
Our aim is to provide the necessary theoretical results, to
next-to-leading order (NLO) accuracy in the strong coupling $\alpha_s$,
for using \gjj-jet production to estimate \Zjj-jet production.
We use the same software
tools as in our previous studies of \Wjn- and \Zjn-jet
production~\cite{W3j,W4j,BlackHatZ3jet,Wpolarization}, the \BlackHat{}
library~\cite{BlackHatI,BlackHatII} along with \AMEGIC{}~\cite{Amegic}
within the \SHERPA{}~\cite{Sherpa} framework, to perform the
leading-order (LO) and NLO calculations.  We also present results for
a parton-shower calculation matched to fixed-order LO matrix elements
(ME+PS)~\cite{HoechePhoton}, also within the \SHERPA{}~\cite{Sherpa}
framework.  A key issue is the theoretical uncertainty in the
conversion from $\gamma$ to $Z$.  We use the difference between the
ME+PS results and the NLO predictions to estimate the uncertainties
for ratios.  (The common variation of factorization and renormalization
scales in the numerator and denominator of these ratios produces quite
small shifts in the ratios, which are likely to underestimate the 
uncertainties substantially.)  The results
presented here are being used by the CMS collaboration in their study
of missing energy in association with three jets~\cite{CMSMET}.  While
our present study is for missing energy in association with two jets,
we do not expect much difference in the theoretical uncertainties
between the two- and three-jet cases.

Photon isolation is essential for rejecting copious hadronic
backgrounds.  The type of photon isolation criterion affects the
theoretical description of the photon production process.  In the
past, various types of isolation cones have been used, which limit the
amount of hadronic energy near the photon candidate.  Fixed isolation
cones generally limit the total amount of energy in a cone, while the
one proposed by Frixione~\cite{Frixione} consists of a set of energy
constraints that become increasingly restrictive the closer one gets
to the photon.  The Frixione cone is theoretically attractive because
it eliminates contributions from long-distance collinear fragmentation
of partons into photons.  Although there is a perturbative
factorization available for other types of cones, the required photon
fragmentation functions~\cite{PhotonFragmentation}
 (non-perturbative functions analogous to the
parton distribution functions) are not known particularly precisely.
From an experimental point of view, some hadronic energy must be
allowed everywhere within the cone in order to cope with the
underlying event and with event pile-up.  In our study, we adopt a
modified cone criterion of the Frixione type. The CMS collaboration
has recently published~\cite{CMSInclusivePhoton} a measurement of the
photon spectrum. We compare the two types of cone isolation to the
data, and show that with our choice of parameters the difference
between them is small for the kinematic region studied by CMS.

In the following section we describe the details of our calculation
and discuss the photon isolation criterion.
Section~\ref{InclusivePhotonSection} presents our cross-check using
isolated prompt-photon production.  Section~\ref{Cuts} discusses
the cuts we use.
In section~\ref{LHCsec} we present the ratios
of \Zjj-jet to \gjj-jet rates for a variety of distributions.
Our conclusions and outlook follow.

\section{The Calculation}

In this section we discuss our calculational setup.
At NLO, we follow the same
basic setup used in refs.~\cite{BlackHatZ3jet,W3j,W4j}
while for the ME+PS study we use the setup of
ref.~\cite{HoechePhoton}.

\subsection{Matrix Elements and Integration}

We compute the cross sections at NLO using the Catani--Seymour dipole
subtraction method~\cite{CS}.  This method requires the combination of
several contributions: the LO term; virtual corrections from the
interference of tree-level and one-loop amplitudes;
the real-emission corrections with dipole subtraction terms;
and the singular phase-space integrals of the dipole terms.

\begin{figure}[t]
\begin{center}
 \subfigure[]{\label{fig:Zdiagram}\includegraphics[clip,scale=0.45]{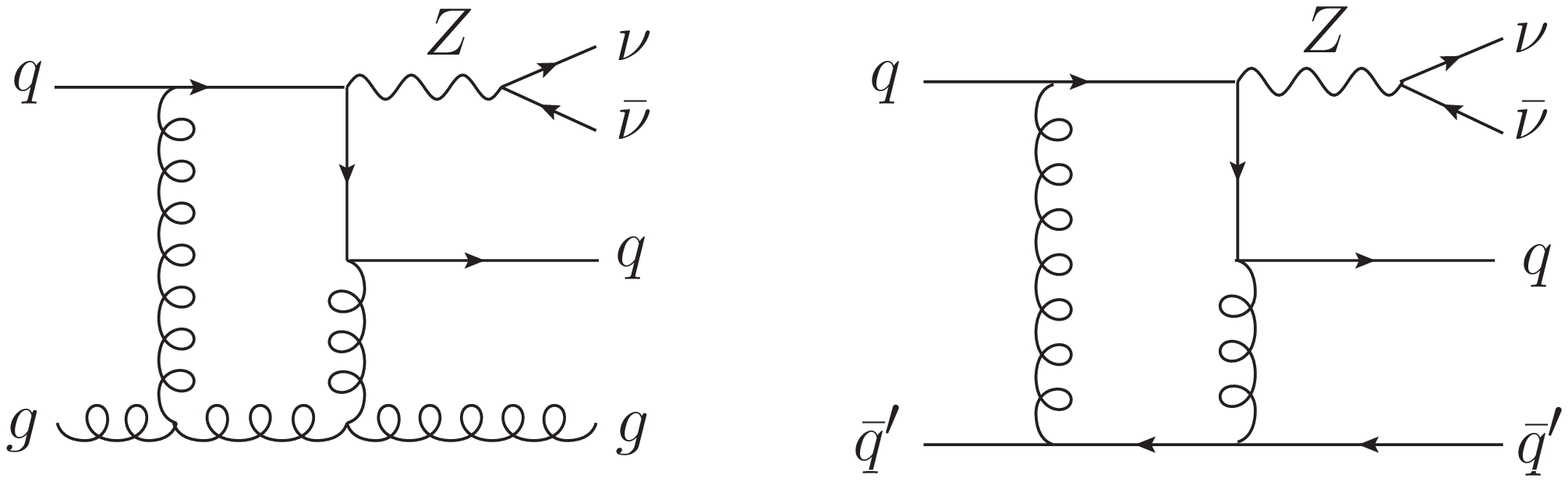}}\\
 \subfigure[]{\label{fig:Gamdiagrams}\includegraphics[clip,scale=0.45]{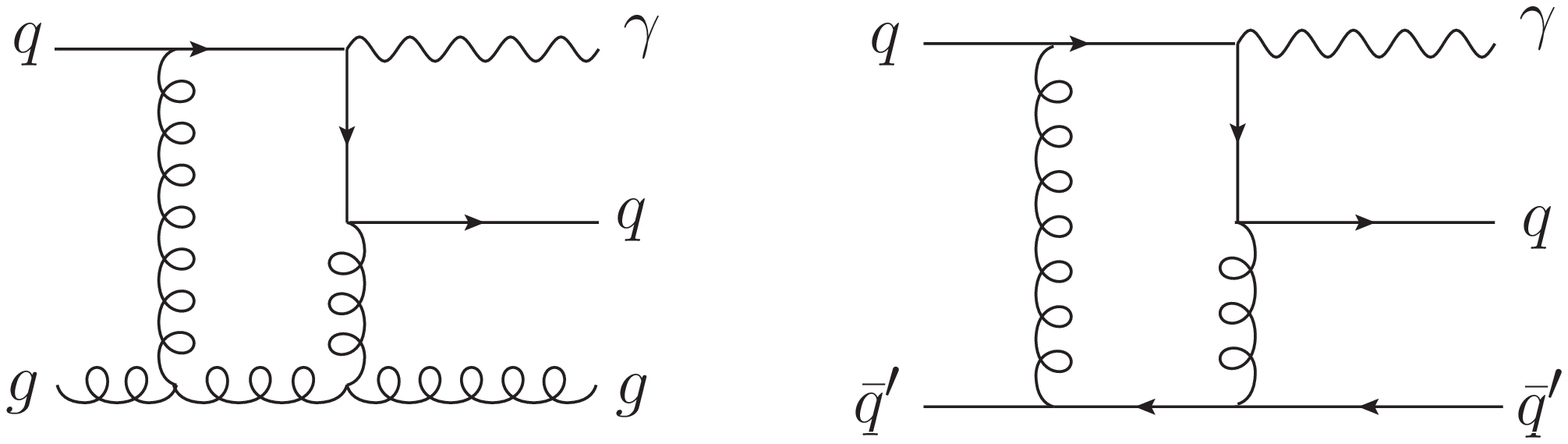}}
\end{center}
\caption{Sample virtual diagrams needed for (a) 
$pp \rightarrow Z(\rightarrow \nu\nub)+2$-jet production and
for (b) $pp \rightarrow \gamma+2$-jet production.}
\label{fig:VirtualExamples}
\end{figure}

We evaluate the required one-loop amplitudes using the \BlackHat{}
program library~\cite{BlackHatI}.  For the processes we are studying,
we need the one-loop corrections to the following partonic processes,
\begin{eqnarray}
 && q \qb g g\rightarrow Z(\rightarrow \nu\nub)\ \hbox{or}\ \gamma\,, 
\nonumber \\
 && q \qb q' \qb'\rightarrow Z(\rightarrow \nu\nub)\ \hbox{or}\ \gamma\,,
\label{VirtualProcesses}
\end{eqnarray}
where two of the four partons should be crossed into the final state, and the
$Z$ decay to neutrinos is folded in.  Some sample diagrams for
these processes are shown in \fig{fig:VirtualExamples}, illustrating
the similarity of the $Z$ and $\gamma$ cases.

For the $Z$ processes, the \BlackHat{} code 
library~\cite{BlackHatI,BlackHatII}
implements analytic one-loop amplitudes from refs.~\cite{Zqqgg,Zqqqq}.
(See also ref.~\cite{OtherZpppp}.)  The photonic amplitudes are
implemented using analytic formul\ae{} in a similar way; they can be
obtained as appropriate sums of color-ordered primitive amplitudes for the
two-quark three-gluon~\cite{qqggg} and four-quark one-gluon
processes~\cite{Kunsztqqqqg}, as explained in ref.~\cite{SignerPhoton}
and in the fourth appendix of ref.~\cite{qqggg}.
This conversion is possible because the trace-based color decomposition
does not distinguish between different generators of $U(N)$.
Setting one of the generators equal to the identity matrix, and collecting
the coefficients of the identical remaining color structures, generates
the photonic amplitudes.  This procedure removes the unwanted three-boson
couplings of the photon present in each primitive amplitude.
We omit the process $gg\to gg\gamma$ as it only contributes to \gjj-jet
production at next-to-next-to-leading order; for the kinematics of
interest here the gluon luminosity is not large enough for this process to
be important.

The NLO result also requires real-emission corrections to the LO process,
which arise from tree-level amplitudes with one additional parton.  
We use the program~\AMEGIC~\cite{Amegic} to compute these
contributions, along with the Catani-Seymour dipole subtraction
terms~\cite{CS} and their integrals over phase space.
The \SHERPA{} framework~\cite{Sherpa} includes
\AMEGIC{} and tools to analyze the results and construct a wide
variety of distributions.  We have previously
validated~\cite{W3j} the \BlackHat+\SHERPA{} framework for
$W + (n\le2)$ jets against the MCFM code~\cite{MCFM}.

We use \SHERPA{} in a second mode, to provide a parton-shower
prediction matched to tree-level matrix elements (ME+PS), also known
as matrix-element-plus-truncated-shower.  (Our parton-shower results
do not include hadronization effects, but remain at the parton level.)
The ME+PS event samples are produced following
ref.~\cite{HoechePhoton}, using the \COMIX{} matrix-element
generator~\cite{Comix}.  This method combines LO hard matrix elements
together with parton showers, which resum logarithmic corrections due
to Bremsstrahlung effects.  The parton shower employed to this end in
\SHERPA{}~\cite{CSShower} is based on Catani-Seymour dipole
factorization~\cite{CS}. In contrast to earlier parton showers, the
procedure inherently respects QCD soft color coherence.  The procedure
allows the unambiguous identification of a recoil partner for partons
that are shifted off mass-shell in the splitting process (the
``mother'' partons). This eliminates one of the major sources of
uncertainty in earlier schemes for parton evolution.  As the
observables presented below should be relatively insensitive to
hadronization effects, ME+PS results are presented at the parton
level.  We match to matrix elements containing up to three final-state
partons, and use 15~GeV for the merging cut.  (Further details may be
found in ref.~\cite{PhotonShower}.)

We work to leading order in the electroweak coupling. The
$Z$-boson couplings are as given in ref.~\cite{BlackHatZ3jet}.
In particular, the $\nu\nub$ invariant mass is distributed in a relativistic
Breit-Wigner of width $\Gamma_Z = 2.49$~GeV about the $Z$ boson mass
of $91.1876$~GeV. These values, along with those of
$\alpha_{\rm QED}(M_Z) = 1/128.802$
and $\sin^2\theta_W = 0.230$ lead to a branching ratio for the neutrino
mode in $Z$ decay of ${\rm Br}(Z\to\nu\nub) = 0.2007$.
We use MSTW2008 leading and next-to-leading order parton distribution
functions, with the QCD coupling
$\alpha_s$ chosen appropriately in each case. Our LHC results are for
a center-of-mass energy of $7$ TeV.

The question of what scale should be chosen for the electromagnetic
coupling in a prompt-photon calculation is a subtle one.  Strictly speaking,
this scale is undetermined in a leading-order computation; and although
we are working to NLO in the QCD coupling, we are only working to LO
in the electromagnetic one.  From a practical point of view, however,
it makes a difference whether we choose a scale of order the photon
transverse momentum (and hence
$\alpha_{\rm EM}\approx \alpha_{\rm EM}(M_Z) \approx 1/128$), or
the zero-momentum-squared value (that is, $\alpha_{\rm EM}(0) = 1/137.036$).

Heuristically, we see that the sequence of fermion bubbles on the photon
forms a gauge-invariant set.  Moreover, because of QED Ward identities,
this is the only set of diagrams which controls the coupling
renormalization.  In the idealized situation we are considering,
the photon is always resolved, {\it i.e.}~it is not allowed to
split into low-mass lepton pairs.  Then summing the bubbles leads to a
zero-momentum-squared coupling for the emission of the hard photon.
Other types of QED effects, such as emission of additional hard photons,
change the kinematics and cannot be absorbed into a running coupling.
This argument is confirmed by the analyses in
refs.~\cite{KniehlLonnblad,MarcianoAlpha}.
We therefore take
the electromagnetic coupling for the prompt-photon computation 
to be the zero-momentum-squared value, $\alpha_{\rm EM}(0) = 1/137.036$,
and {\it not\/} the running value at typical collider energies.

\subsection{Infrared Safety and Photon Isolation}
\label{PhotonIsolationSection}

An observable in perturbative QCD is infrared- and collinear-safe if
it is unaltered under the emission of an arbitrarily soft gluon, or
the splitting of a colored parton into a pair of colored partons
(whether in the initial or final state).
We follow ATLAS and CMS in using the infrared-safe
anti-$k_T$ jet algorithm~\cite{antikT}.

For \gjj-jet production, there is an additional infrared issue beyond
the infrared safety of the jet algorithm.  Because photons can arise
from $\pi^0$ decay and other hadronic sources, it is
essential from an experimental point of view to insist that they be
isolated from jets if we want to study photons that originate in
short-distance physics (`prompt').  On the other hand, too strict a
photon isolation criterion --- such as requiring no hadrons or no
hadronic energy in a cone around the photon, or using tracking alone
to determine isolation --- would be infrared-unsafe, and prevent the
use of perturbative QCD as a theoretical tool.  In order to navigate
between these two competing requirements, experimental collaborations
typically use a weighted isolation criterion, imposing a limit on the
hadronic energy fraction in a cone around the photon, or simply on
the total hadronic energy in the cone.

In their recent measurement of the inclusive isolated prompt-photon 
spectrum~\cite{CMSInclusivePhoton}, the CMS collaboration
required photon candidates to satisfy a set of requirements
on nearby energy deposits measured via tracking, and
in the electromagnetic and hadronic calorimeters.  In their theoretical
modeling using P{\sc ythia}, they required photons to have
less than 5~GeV of summed $p_T$ within an isolation cone of radius $R=0.4$,
where $R_{i \gamma} = \sqrt{
  (\eta_i - \eta_\gamma)^2 + (\phi_i - \phi_\gamma)^2}$.  We
will adopt this criterion as our `reference' standard cone isolation
in our discussion of the isolated prompt-photon spectrum.

While such a criterion is infrared-safe with respect to the strong
interactions --- emission of a soft gluon, or a colored parton
splitting into two colored partons --- it is {\it not} collinear-safe
with respect to QED: the cross-section receives contributions from
collinear radiation of photons off massless quarks.  In a theoretical
description, this singularity has to be factorized, and absorbed into
parton-to-photon fragmentation functions~\cite{PhotonFragmentation}, 
whose computation from first
principles would require knowledge of non-perturbative physics.  The
factorization and the non-perturbative functions are the final-state
analogs of the parton distribution functions.  In practice, these
functions are extracted from fits to experimental data, although these
fits are not nearly as precise as those for the parton distributions,
nor are sets surrounding a central fit available to estimate errors.
(As is true for the parton distributions, the evolution of the
fragmentation function with the hard scale can be determined in
perturbative QCD.)

However, the isolation criterion given above is not the only possible
one.  Frixione~\cite{Frixione} proposed a modified isolation
requirement which suppresses the collinear region of the phase space and 
thereby eliminates the need for a fragmentation-function contribution.
We follow this proposal, requiring that each parton $i$
within a distance $R_{i\gamma}$ of the photon obey
\begin{equation}
 \label{iso} \sum\limits_i E_{iT}\, \Theta \left(\delta -
R_{i\gamma}\right) \leq {\mathcal{H}} (\delta) \,,
\end{equation}
for all $\delta \leq \delta_0$, in a cone of fixed half-angle
$\delta_0$ around the photon axis. The restricting function
${\mathcal{H}} (\delta)$ is chosen such that it vanishes as $\delta
\rightarrow 0$ and thus suppresses collinear configurations, but
allows soft radiation arbitrarily close to the photon. We adopt 
\be
{\mathcal{H}} (\delta) = E_{T}^\gamma \, \epsilon 
\left( \frac{ 1 - \cos \delta }{1 - \cos \delta_0 } \right)^n \, ,
\label{photoniso}
\ee
where $E_T^\gamma$ is the photon transverse energy.

An experimental analysis cannot adopt the Frixione prescription
precisely, because of the finite resolution of detectors.  One
can imagine using a discretized version of it; this would still require
a fragmentation contribution (corresponding to the innermost step
of the discretized cone), but it would presumably have a much smaller
one than the standard cone.  One could
imagine adjusting the parameters $\epsilon$, $\delta_0$ and $n$ so
as to minimize that contribution.
A preliminary study of the comparison
between the Frixione prescription and a discretized version may be
found in ref.~\cite{LesHouchesDiscretizedFrixioneStudy}.  

In our study, we will use the Frixione cone, with $\epsilon = 0.025$,
$\delta_0 = 0.3$ and $n = 2$.  In practice, we find that our
predictions are only weakly sensitive to these parameters.
In the next section, we will compare
our predictions using these parameters to predictions made using
a standard cone for the isolated prompt-photon spectrum~\cite{CMSInclusivePhoton} 
measured by CMS.  As we shall see, the differences between the Frixione
and standard-cone isolation prescriptions are not large, and are
quite small in the large-$p_T^\gamma$ region that is of primary
interest in the present study.  We therefore conclude that it is reasonable
to use the Frixione isolation to model the $Z$-to-photon ratio in association
with two jets for CMS's analysis.

In the definition of the cross section for $\gamma+m$ jets, at least
$m$ jets must lie outside the isolation cone, while we allow any
number of additional jets $m'$ to fall inside the cone of radius
$\delta_0 = 0.3$. In order to obtain the desired observable we apply
the jet-finding algorithm to all partons except the photon in the
event, to obtain $m$ jets outside the isolation cone and $m'$ jets
inside the cone.  Then we apply the cuts on transverse momenta and
rapidity of the $m$ jets.  This procedure is infrared safe because the
jet-finding algorithm is applied everywhere and not just outside the
isolation cone. In practice, jets appear very rarely inside the photon
isolation cone, because of the combination of the jet $p_T$ cut and
the small energy fraction we use.  We will list the other cuts we
apply in a later section.

\section{Isolated Prompt-Photon Production}
\label{InclusivePhotonSection}

Recently, CMS has published a measurement~\cite{CMSInclusivePhoton} of
the inclusive isolated prompt-photon spectrum based on a data sample of
2.9~pb${}^{-1}$ from early running of the LHC.  They compared the
measurement with NLO predictions from the publicly-available
\jetphox{} code~\cite{JetPhoX}.  These predictions make use of a
standard photon isolation cone as described in the previous section,
and include fragmentation contributions.
(The \jetphox{} code implements a Frixione isolation cone as well
as a standard isolation cone, although the former implementation is
not used in this study.)  We can use the CMS study to
assess the expected differences between the use of standard-cone
isolation and a Frixione isolation for the photon.  To do so, we have
made use of an NLO code, due to Gordon and Vogelsang~\cite{Vogelsang}.
It is a semi-analytic code,
relying on the narrow-cone approximation. We also used \BlackHat{} in
conjunction with \SHERPA{} as a cross check on the NLO Frixione
isolation result.  (In all these calculations, $\alpha_{\rm EM}$ was
set to $1/137$, for the reasons noted above.)

\begin{figure}[t]
\begin{center}
\includegraphics[clip,scale=0.55]{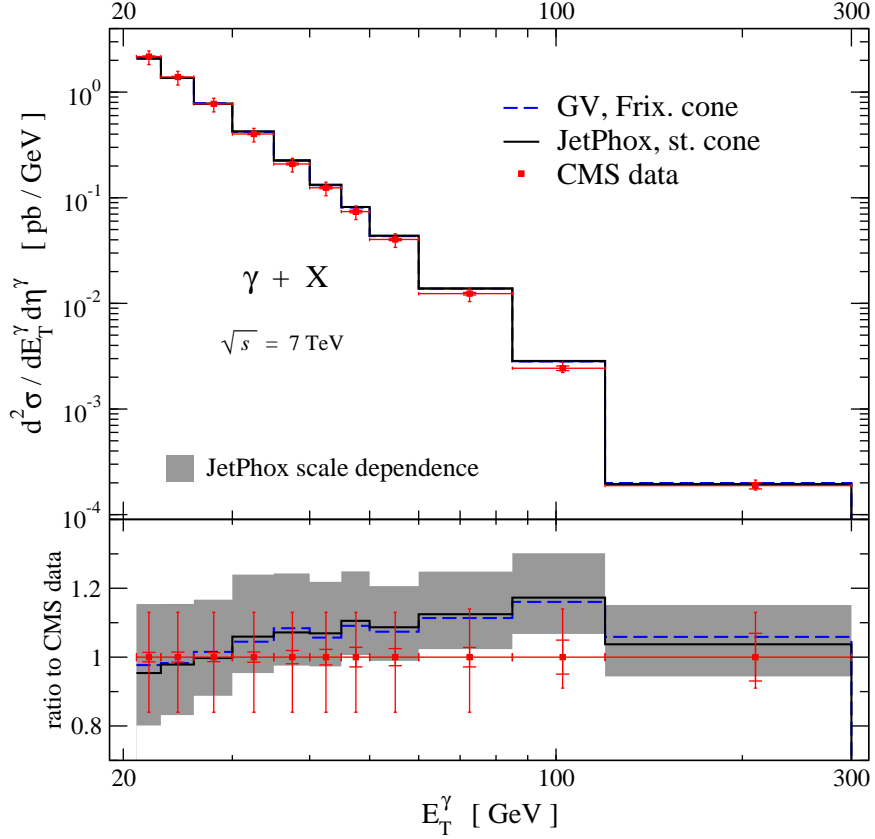}
\end{center}
\caption{A comparison of different NLO theoretical predictions for $\gamma+X$
production at the LHC at 7~TeV.  The CMS data points~~\cite{CMSInclusivePhoton}
 are shown (red) with
combined experimental uncertainties; the prediction using 
using the Gordon-Vogelsang (GV) code and 
  the Frixione isolation criterion is given by the dashed (blue) line;
the \jetphox{} prediction using
  a standard cone isolation is given by the solid (black) line.  
  The lower panel shows the two theoretical
  predictions normalized to
  the CMS data along with the scale dependence (shaded gray) as determined using
  \jetphox.  }
\label{fig:gammaIncsv_CMS_jetphox_GV}
\end{figure}

\begin{figure}[t]
\begin{center}
\includegraphics[clip,scale=0.55]{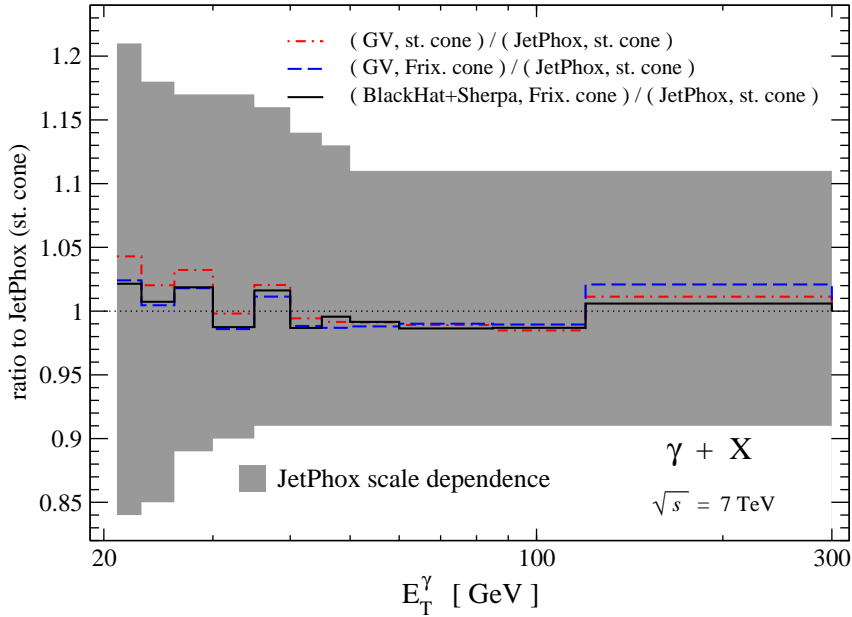}
\end{center}
\caption{A comparison of four different NLO predictions for
$\gamma + X$ production at the LHC at 7~TeV.  All predictions
are normalized to the \jetphox{} predictions shown 
in \fig{fig:gammaIncsv_CMS_jetphox_GV}.  The prediction for
a standard cone isolation using the Gordon-Vogelsang (GV) code is
given by the dot-dashed (red) line; that for the Frixione isolation
criterion using the GV code, by the dashed (blue) line; and
that for the Frixione isolation using the BlackHat+Sherpa code, by
the solid (black) line.   The \jetphox{} scale dependence band is
shown in gray.
}
\label{fig:gammaIncsv_CMS_theory}
\end{figure}

The results of the comparison are shown in
\figs{fig:gammaIncsv_CMS_jetphox_GV}{fig:gammaIncsv_CMS_theory}
 as a function of the photon $p_T$.
\Fig{fig:gammaIncsv_CMS_jetphox_GV} shows
the CMS measurement, with its statistical and systematic uncertainties,
along with the NLO predictions for a fixed-cone isolation as given
using~\jetphox{} in ref.~\cite{CMSInclusivePhoton}, 
and for the Frixione isolation criterion computed
using the Gordon--Vogelsang code.  The scale variation band
shown was computed using \jetphox{}.  \Fig{fig:gammaIncsv_CMS_theory}
compares several different NLO predictions to the
\jetphox{} prediction as given in ref.~\cite{CMSInclusivePhoton}:
 fixed-cone and Frixione isolation ones, using the Gordon--Vogelsang code,
and one for the Frixione isolation, using \BlackHat{} with \SHERPA.
All the predictions here use
the CT10 parton distributions~\cite{CT10}, which have $\alpha_s(M_Z) =
0.1179$.  We see that the fixed-cone prediction from the
Gordon--Vogelsang code, while not identical to the \jetphox{} one, is
in excellent agreement, well within the experimental systematic
uncertainties.  In addition,
the predictions for the Frixione-style isolation are also quite close,
within 2\% at low $p_T^\gamma$, and within 1\% at high $p_T^\gamma$.
It is the large-$p_T$ region that is relevant to the primary study we
perform in this paper.  We expect that the smallness of this
difference will carry over to processes with multiple jets in addition
to the photon present here.  We will include this difference in our
overall uncertainty estimate for the $Z$-to-photon ratios.  (See
ref.~\cite{FrixioneVogelsangPolarized} for a preliminary
comparison between the two isolation criteria in the context of
polarized $pp$ scattering.)
\Fig{fig:gammaIncsv_CMS_theory} also verifies the agreement in results from the 
Gordon--Vogelsang code for the Frixione-style isolation with
those from \BlackHat{}+\SHERPA{}  to
within 1\%, uniformly in $p_T^\gamma$.  In this particular \BlackHat{}+\SHERPA{}
calculation, we impose no jet requirements, and impose
a fixed maximum hadronic 
energy of 5~GeV in the isolation definition~(\ref{photoniso})
in place of $E_{T}^\gamma\,\epsilon$ with $\epsilon=0.025$
as done elsewhere in our \gjj-jet study.  (The difference 
in cross sections between using a fixed energy of 5~GeV and
the energy fraction used in our study is less
than 1\% for $p_T^\gamma>50$~GeV.)

The rest of our study makes use of the MSTW08 parton distributions. Predictions
with this set, which has a larger $\alpha_s(M_Z)$ of $0.1201$, are
4\% higher than those shown in \fig{fig:gammaIncsv_CMS_jetphox_GV},
fairly uniformly in $p_T^\gamma$; however, this difference is expected
to become much smaller when taking the ratio of \Zjj- to \gjj-jet
quantities.  

\section{Photon-to-$Z$ Ratios at the LHC}
\label{Cuts}

\def\firstJ{{\rm 1}^{\it st}{\, \rm jet}}
\def\secondJ{{\rm 2}^{\it nd}{\, \rm jet}}
Our focus in this paper is on using distributions measured for \gjj-jet
production to predict similar distributions assembled from 
missing~$E_T$\,+\,2-jet events.
Ideally, we would compute the ratio of fully-differential cross sections,
\be
{{d\sigma^Z/dA}\over {d\sigma^\gamma/dA}} \,,
\ee
where
\be
{d\sigma^V\over dA} \equiv {d\sigma^{V+2{\rm jets}}\over dE_T^{\firstJ}
 d\eta^{\firstJ}  d\phi_{\vphantom{T}}^{\firstJ}
dE_T^{\secondJ} d\eta_{\vphantom{T}}^{\secondJ} 
 d\phi_{\vphantom{T}}^{\secondJ}}
\,.
\ee
This information could be used experimentally by adjusting the weight
of each \gjj{} jet event, removing the photon, and then reweighting
the sample to obtain unit-weight events.  This would provide a sample
of estimated \Zjj{} jet events with the $Z$ decaying invisibly.

With sufficient statistics we could effectively construct such a
quantity using $n$-tuples, tracking the matrix-element weights and
parton momenta.  However, besides the practical issue of carrying out such 
a procedure at NLO, we would have
no way to usefully present such a quantity in two-dimensional
form. Accordingly, we will study the ratios of a variety of
singly-differential distributions.  This can be thought of as a
projected version of the above ratio, with a projection onto one of
the axes in the complete differential cross section.

We use the anti-$k_T$ jet algorithm~\cite{antikT} with clustering parameter $R =
0.5$, where $R = \sqrt{(\Delta y)^2+(\Delta\phi)^2}$ as usual
in terms of rapidity and azimuthal angle.

Our Monte Carlo set up allows us to study any infrared and
collinear-safe observable.  In the present study, we focus on three
sets of cuts, theoretical parallels to those used by the CMS
collaboration~\cite{CMS-photon-note} and intended to be relevant in
different regions of the SUSY parameter space.  In order to display
these cuts, we make use of a special definition of the total
transverse energy, which we label $\HTj$, as the sum of the transverse
energies of all jets with $p_T > 50$~GeV and $|\eta| < 2.5$.   We also define a
vector MET, as the negative of the four-vector sum of all jets with
$p_T > 30$~GeV and $|\eta|<5$.  Each set of cuts that we will consider is
distinguished by different restrictions on the quantities\footnote{MET
  stands for `Missing Transverse Energy'. Although this is standard
  terminology, it is potentially misleading, as the MET is in fact the
  missing transverse momentum.}  $\HTj$ and MET.

  \begin{description}
    \item[Set 1:] {$\HTj > 300$ GeV}, {$|\textrm{MET}| > 250$ GeV},
    \item[Set 2:] {$\HTj > 500$ GeV}, {$|\textrm{MET}| > 150$ GeV},
    \item[Set 3:] {$\HTj > 300$ GeV}, {$|\textrm{MET}| > 150$ GeV}.
  \end{description}

For all sets we insist that the two highest-$p_T$ jets have $p_T$ at
least $50$ GeV and pseudorapidity of at most $|\eta|= 2.5$. These jets
are referred to as `tagging jets'. We note that beyond leading order
there can be other jets in the event.  The separation in $\phi$-space
between each tagging jet and the MET vector is required to satisfy
$\Delta\phi(\textrm{jet},\textrm{MET}) > 0.5$.

In addition to the above cuts, for the \gjj{} jet study only, we impose
photon isolation according to the Frixione~\cite{Frixione}
prescription, with parameters $\epsilon = 0.025$, $\delta_0 = 0.3$ and $n = 2$,
and a minimum $R$-space separation between the MET vector and each
tagging jet of 0.4. The photon is required to have $|\eta| < 2.5$.
We also impose a minimum $p_T$ of 100~GeV on both the photon and the $Z$.
This cut has no effect because of the $|\textrm{MET}|$ cut, but it 
improves the numerical efficiency of the calculation.

The Set 1 cuts can be roughly characterized as the low-$\HTj$ /
high-MET region, whereas Set 2 is the converse, high-$\HTj$ /
low-MET. The reason for studying these two sets is that different
SUSY production mechanisms are expected to lead to signals in
different regions. Broadly speaking, Set 1 is geared towards catching
direct squark decays, while Set 2 is designed for cascades with a
$W$-boson and a softer lightest supersymmetric particle (LSP). Set 3,
which is inclusive of both the others, is a control region.

Our fixed-order results depend on the renormalization and
factorization scales. These scales are unphysical, but necessarily
appear when the perturbative series is truncated at a finite order.  For
fixed-order predictions, it is customary to estimate the error
arising from omission of higher-order terms by varying these scales
around some central value. The size of the resulting band is a useful
diagnostic for those situations where fixed-order perturbation theory
breaks down. The central value should be a typical hard scale
in the process, to minimize the impact of potentially large
logarithms. We choose the dynamical scale $\mu = H_T'/2$ for this central
value, where $H_T'$ is defined as
\begin{equation}
H_T' = \sum_i E_T^i + E_T(Z,\gamma) \,,
\end{equation}
where $i$ runs over the partons and $E_T \equiv \sqrt{ M^2+p_T^2}$. A
common method for estimating the error on a cross section is by
varying the common scale up and down by a factor of two. We do so by
evaluating the cross sections at five scales: $\mu/2, \mu/\sqrt2, \mu,
\sqrt2\mu, 2\mu$.  As we will discuss below, this procedure is
expected to greatly underestimate uncertainties when applied to a
ratio of cross sections.


\section{LHC Predictions}
\label{LHCsec}

\begin{table}[t]
  \begin{center}
\begin{tabular}{|c||c|c|c|}
\hline
 process  & LO  &ME+PS  & NLO  \\
\hline
{\tt $Z+2j$}         & $0.521(0.001)^{+0.180}_{-0.125}$ & $0.416(0.004)$ & $0.560(0.002)^{+0.012}_{-0.042}$ \\
\hline
{\tt $\gamma + 2j$}  & $2.087(0.005)^{+0.716}_{-0.494}$ & $1.943(0.027)$ & $2.448(0.008)^{+0.142}_{-0.225}$ \\
\hline
$Z/\gamma$ ratio  & $0.250$ & $0.214$ & $0.229$ \\
\hline
\end{tabular}
\caption{Set 1 cross sections for $Z$ and $\gamma$ production in association with
  two jets, using the anti-$k_T$ jet algorithm. The numbers in
  parentheses are Monte Carlo statistical errors, while the upper and
  lower limits represent scale dependence. See the text for a
  discussion of the errors on the ratio.\label{Set1XS} }
  \end{center}
\end{table}

\begin{table}[t]
  \begin{center}
\begin{tabular}{|c||c|c|c|}
\hline
 process  & LO &ME+PS  &  NLO  \\
\hline
{\tt $Z + 2j$}       & $0.205(0.001)^{+0.073}_{-0.050}$ & $0.238(0.005)$ & $0.277(0.002)^{+0.032}_{-0.033}$ \\
\hline
{\tt $\gamma + 2j$}  & $0.952(0.004)^{+0.333}_{-0.230}$ & $1.132(0.010)$ & $1.374(0.008)^{+0.268}_{-0.148}$ \\
\hline
$Z/\gamma$ ratio  & $0.215$ &  $0.211 $ & $0.201$ \\
\hline
\end{tabular}
\caption{As in Table 1, but for the Set 2 cuts.}\label{Set2XS}
  \end{center}
\end{table}

\begin{table}[t]
  \begin{center}
\begin{tabular}{|c||c|c|c|}
\hline
 process  & LO & ME+PS & NLO \\
\hline
{\tt $Z+2j$}       & $1.255(0.002)^{+0.429}_{-0.296}$ & $1.174(0.011)$  & $1.476(0.007)^{+0.090}_{-0.136}$ \\
\hline
{\tt $\gamma+2j$}  & $5.854(0.011)^{+1.975}_{-1.361}$ & $6.075(0.056)$  & $7.601(0.019)^{+0.754}_{-0.826}$ \\
\hline
$Z/\gamma$ ratio          & $0.215              $ & $0.194$  & $0.195$ \\

\hline
\end{tabular}
\caption{As in Table 1, but for the Set 3 cuts.\label{Set3XS} }
  \end{center}
\end{table}

In this section we present total cross sections and distributions for
\gjj-jet and \Zjj-jet production at the LHC running
at $7$ TeV.  We present results
for each of the three sets of cuts. In the \Zjj{} jet
study we fold in the $Z$ boson decay into neutrinos, which comprise the
missing energy. 
The branching ratio for the $Z$ to decay to neutrinos is largely responsible
for the \gjj{} jets cross section being about a
factor of five larger than for $Z(\to\nu\nub)\,\!+\,2$ jets.
This ratio is clearly visible in our figures, and is of course the underlying
motivation for this study. We will also discuss the error to be assigned to our
predictions.

In Tables~\ref{Set1XS}, \ref{Set2XS} and \ref{Set3XS} we give the
total cross section for the cuts outlined in the previous
section. Each table shows three different theoretical
predictions. Fixed-order perturbative results are shown as LO and NLO.
The final states in these cases consist of the vector boson with the
two tagging jets, though there can be an extra jet at NLO. The parton
shower result, labeled ``ME+PS'', is a tree-level matrix element calculation
merged to a parton shower~\cite{HoechePhoton}, as
summarized above. Here the final state can contain many jets, though
virtual corrections are not taken into account.  The LO predictions
are the least reliable of the three and are shown only for reference
purposes.

With the cuts of Sets 1 and~2, the corrections from LO to NLO lead
to an increase in the total cross sections of up to $50 \%$. The
corrections for the control region, given by the Set 3 cuts, are much
more modest.  The ME+PS and NLO results do not agree well for the
cross sections.  However, when one takes the ratio of $Z$ and $\gamma$
cross sections, the two predictions agree to within better than 10\%. 
This behavior is not surprising: typically, overall normalizations can be
somewhat off in ME+PS calculations, while ratios tend to behave much better. 

We have found that in the ratio the LO scale variation cancels nearly
completely, if we vary the scale in a correlated way in the \Zjj-jet
and \gjj-jet predictions.  In the NLO case the scale variation is a
bit larger but also very small.  This nearly complete cancellation of
the scale variation cannot be interpreted as a small theoretical
uncertainty.  The closeness of the NLO and ME+PS ratios is a much
better indication that the theoretical uncertainties on the individual
cross sections do indeed largely cancel in the ratio.  We do not
include the uncertainty due to the parton distributions in our study, 
but we expect it to largely cancel in the ratio since the $d(x)/u(x)$
ratio feeding into it is from a well-measured region in $x$.

\begin{figure}[t]
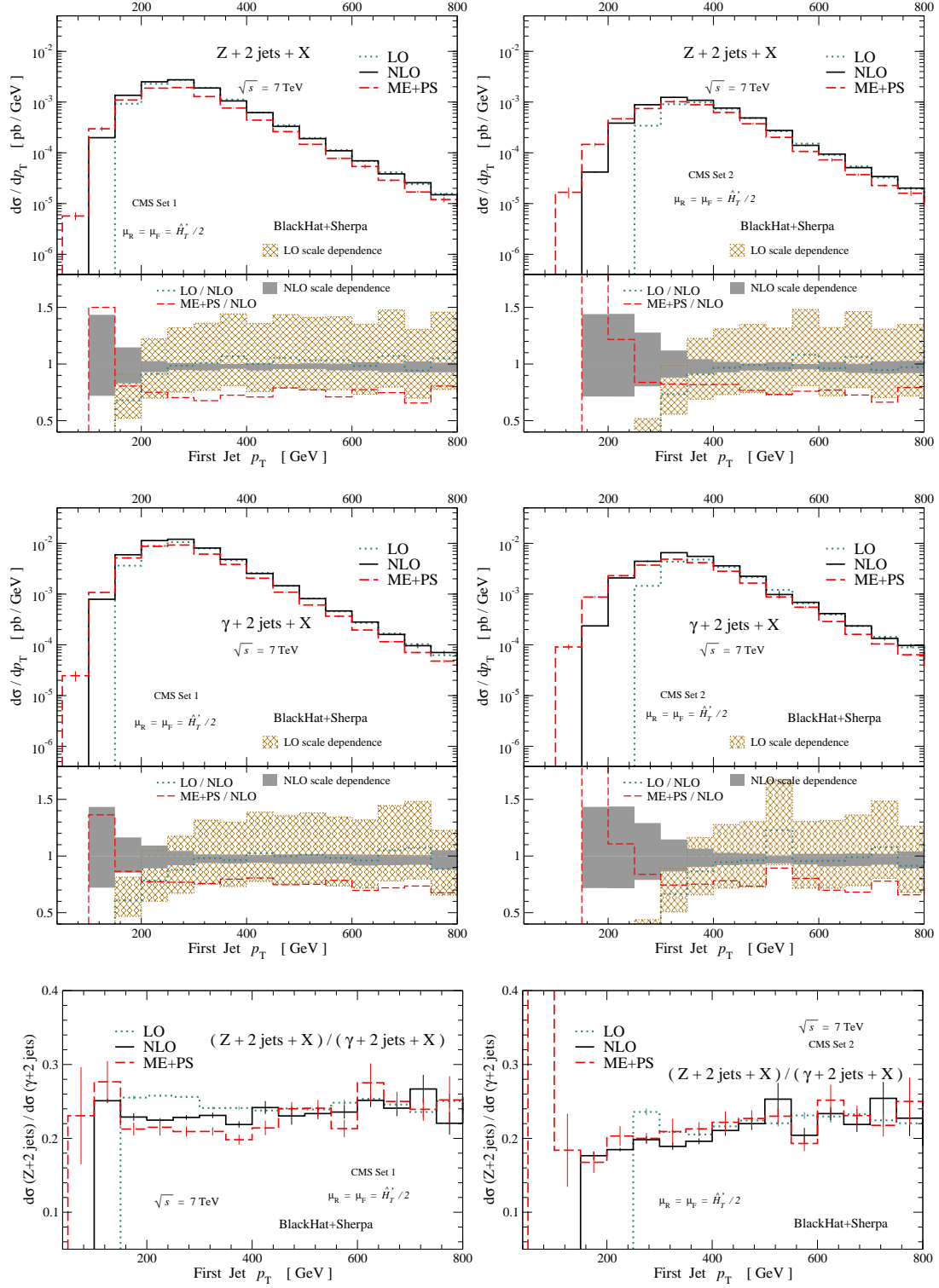

  \begin{center}
    \includegraphics[clip,scale=0.34]{Z2j-CMS-INC_anti-kt-R5-Pt50_jetsHT300MHT250_jet_1_1_pt__A1.eps}
    \includegraphics[clip,scale=0.34]{Z2j-CMS-INC_anti-kt-R5-Pt50_jetsHT500MHT150_jet_1_1_pt__A1.eps} \\
\vskip 10pt
    \includegraphics[clip,scale=0.34]{gm2j-CMS-INC_anti-kt-R5-Pt50_jetsHT300MHT250_jet_1_1_pt__A1.eps}
    \includegraphics[clip,scale=0.34]{gm2j-CMS-INC_anti-kt-R5-Pt50_jetsHT500MHT150_jet_1_1_pt__A1.eps} \\
\vskip 10pt
\hskip 5pt
    \includegraphics[clip,scale=0.34]{Z2j_over_gm2j_NLOMEPS_anti-kt-R5-Pt50_jetsHT300MHT250_jet_1_1_pt__A1.eps}
    \includegraphics[clip,scale=0.34]{Z2j_over_gm2j_NLOMEPS_anti-kt-R5-Pt50_jetsHT500MHT150_jet_1_1_pt__A1.eps}
  \end{center}
  \caption{\baselineskip 14pt%
The $p_T$ distribution of the first jet. The left column shows
distributions for the Set 1 cuts, and the right column for the Set 2 cuts.
Each column displays the differential cross section for \Zjj-jet
production (top), \gjj-jet production 
(middle), and their ratio (bottom).  In the top and middle
plots, the upper panel shows the LO, NLO, and ME+PS results for the
distribution, and the lower panel shows the ratio to the central
NLO prediction, along with the LO and NLO scale-dependence bands. 
The numerical integration uncertainties
are indicated by thin vertical lines.
}
  \label{fig:Jet1PT}
\end{figure}

\begin{figure}[t]
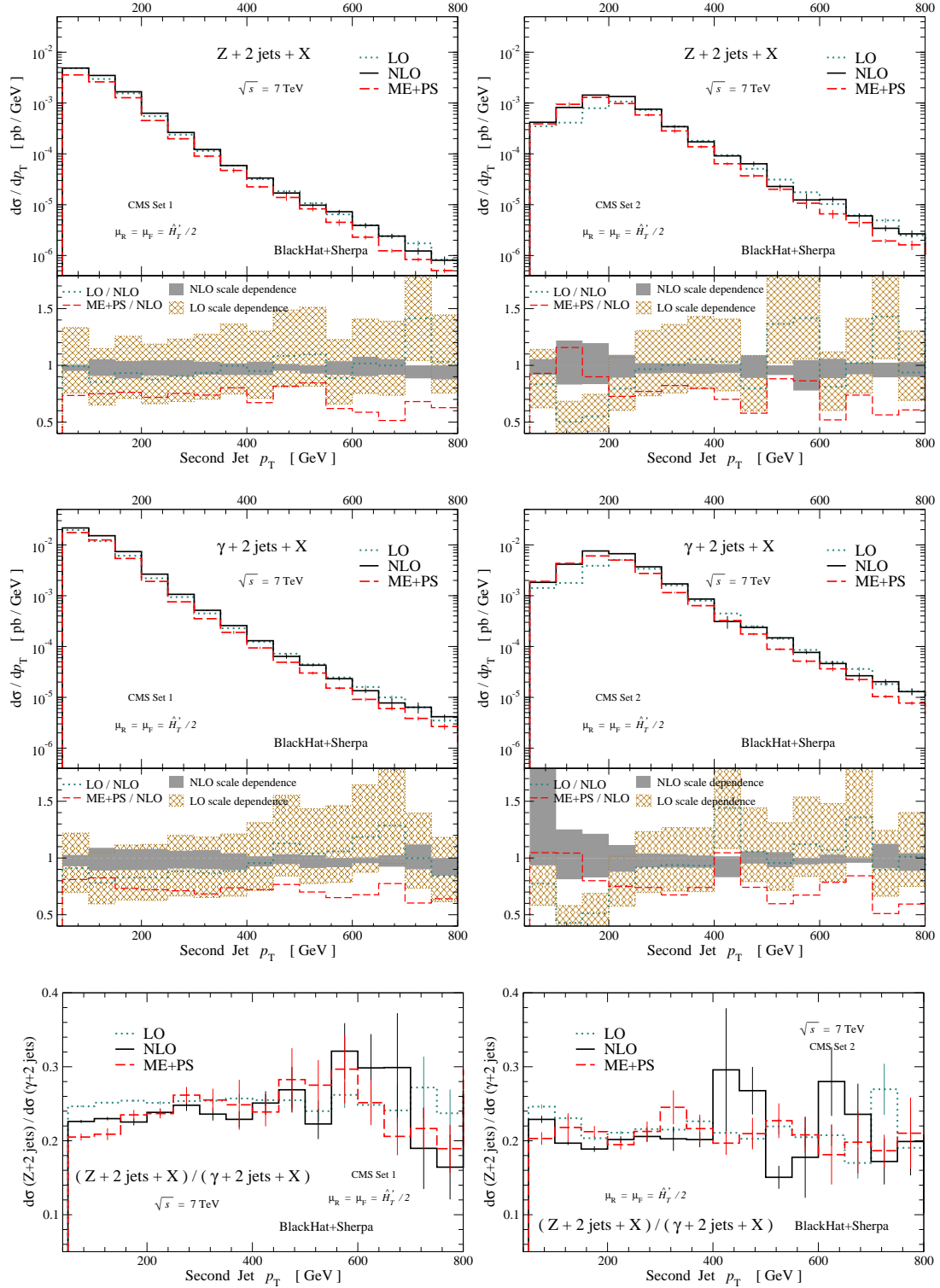

  \begin{center}
    \includegraphics[clip,scale=0.34]{Z2j-CMS-INC_anti-kt-R5-Pt50_jetsHT300MHT250_jet_1_1_pt__A2.eps}
    \includegraphics[clip,scale=0.34]{Z2j-CMS-INC_anti-kt-R5-Pt50_jetsHT500MHT150_jet_1_1_pt__A2.eps} \\
\vskip 10pt
    \includegraphics[clip,scale=0.34]{gm2j-CMS-INC_anti-kt-R5-Pt50_jetsHT300MHT250_jet_1_1_pt__A2.eps}
    \includegraphics[clip,scale=0.34]{gm2j-CMS-INC_anti-kt-R5-Pt50_jetsHT500MHT150_jet_1_1_pt__A2.eps} \\
\vskip 10pt
\hskip 5pt
    \includegraphics[clip,scale=0.34]{Z2j_over_gm2j_NLOMEPS_anti-kt-R5-Pt50_jetsHT300MHT250_jet_1_1_pt__A2.eps}
    \includegraphics[clip,scale=0.34]{Z2j_over_gm2j_NLOMEPS_anti-kt-R5-Pt50_jetsHT500MHT150_jet_1_1_pt__A2.eps}
  \end{center}
  \caption{\baselineskip 14pt%
The $p_T$ distribution of the second jet.
The plots are arranged and the curves are labeled as in \fig{fig:Jet1PT}.
}
  \label{fig:Jet2PT}
\end{figure}

Let us turn next to an examination of five different distributions:
the transverse momentum of the first (largest $p_T$) accompanying jet;
the $p_T$ of the second jet;
the total transverse energy of the jets $\HTj$; and the azimuthal angle distributions
with respect to the MET vector of the first and second
accompanying jets.  These variables are useful for assessing the extent
to which the kinematics of \gjj{}-jet events resembles that of
\Zjj{}-jet events.

The $p_T$ distribution for the leading jet is shown in
\fig{fig:Jet1PT}.  The peak in the distribution is well above the jet
cut, because of the additional cuts on the MET and $\HTj$.
The LO cross section vanishes below 150~GeV for Set 1 and 250~GeV for
Set 2, due to the restricted LO kinematics, which are relieved both
at NLO, from the presence of one additional parton, or in the ME+PS
approximation, which can have several additional partons (including
the parton shower).  Clearly fixed-order LO is inadequate for
describing the $p_T$ distribution for the leading jet. 

The $p_T$ distribution for the second accompanying jet is shown in
\fig{fig:Jet2PT}; we consider only $p_T>50$~GeV.  This distribution is
peaked at the jet cut of $50$~GeV for Set 1, and falls rapidly with
increasing $p_T$.  For Set 2, in contrast, the distribution peaks at a
second-jet $p_T$ of around 180~GeV.  The latter reflects a
compensation for the higher total jet $E_T$ needed to arrive at the
minimum $\HTj$ in this set; a second jet at the lower cut would force
the leading jet too far out onto the tail of its distribution.  The
NLO corrections for Set 1 are approximately 20\%, in line with the
corrections to the total cross section.  These corrections are roughly
flat across the distribution, so that the shape does not change at
NLO.  The NLO corrections for Set 2, in contrast, are significantly
larger, up to 100\% in some bins of second-jet $p_T$, but under 10\%
for $p_T$ above 300~GeV, but below 400~GeV, where the integration
errors become large. The shape of the distribution correspondingly
suffers significant corrections at NLO.  For both sets, however, the
corrections do not distinguish $Z$ and $\gamma$ production, so that
the ratio has NLO corrections of 10--15\%, but flat across all $p_T$s.
The NLO and ME+PS results are likewise in good and uniform agreement
for the ratio, to better than 10\%, again ignoring regions with large
integration errors. As expected, the scale-dependence bands, from
varying the common renormalization and factorization scale,
$\mu_R=\mu_F$, up and down by a factor of two, are narrower at NLO
than at LO.

\begin{figure}[t]
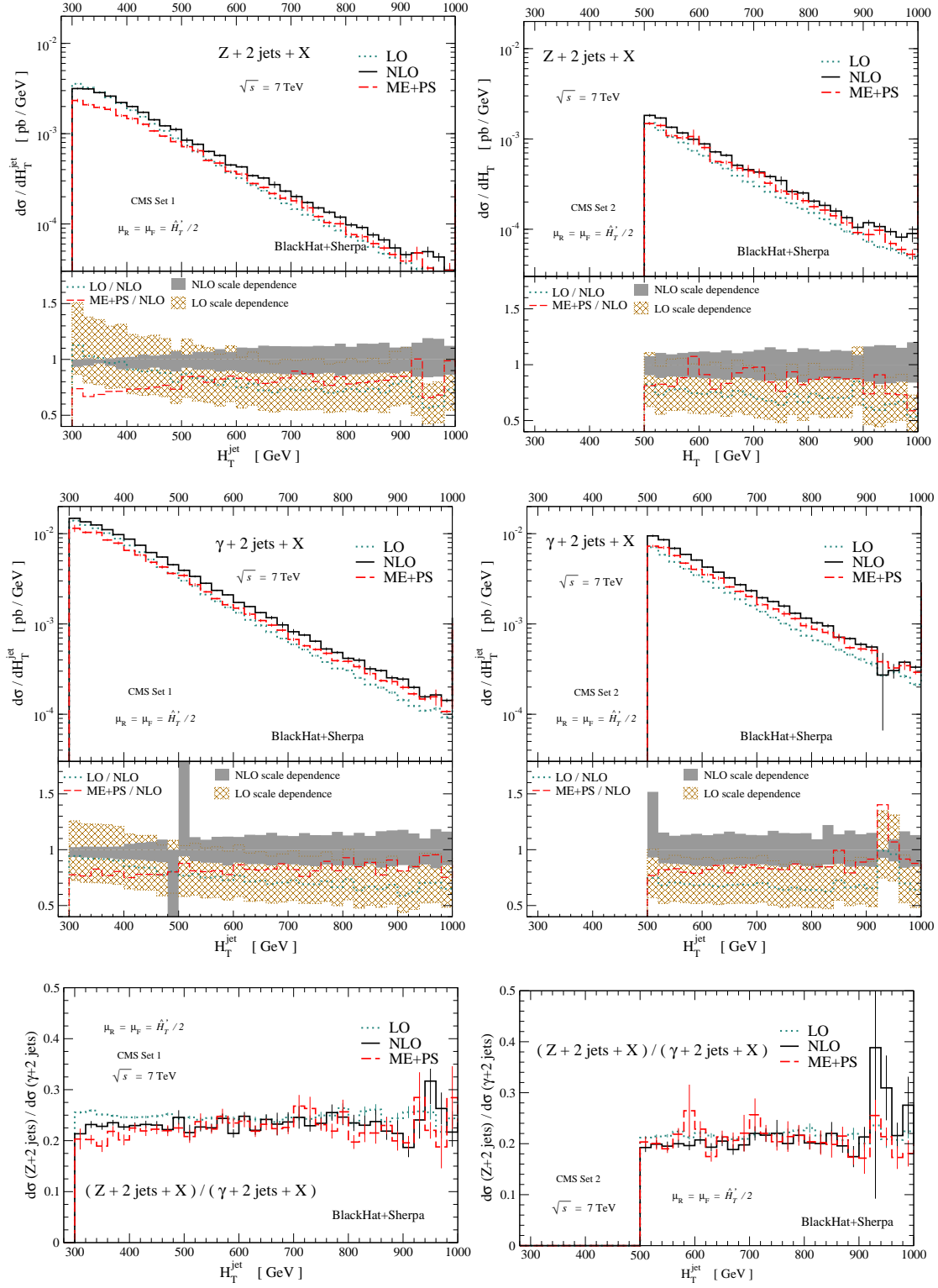

  \begin{center}
    \includegraphics[clip,scale=0.34]{Z2j-CMS-INC_anti-kt-R5-Pt50_jetsHT300MHT250_HT.eps}
    \includegraphics[clip,scale=0.34]{Z2j-CMS-INC_anti-kt-R5-Pt50_jetsHT500MHT150_HT.eps} \\
\vskip 10pt
    \includegraphics[clip,scale=0.34]{gm2j-CMS-INC_anti-kt-R5-Pt50_jetsHT300MHT250_HT.eps}
    \includegraphics[clip,scale=0.34]{gm2j-CMS-INC_anti-kt-R5-Pt50_jetsHT500MHT150_HT.eps} \\
\vskip 10pt
\hskip 5pt
    \includegraphics[clip,scale=0.34]{Z2j_over_gm2j_NLOMEPS_anti-kt-R5-Pt50_jetsHT300MHT250_HT.eps}
    \includegraphics[clip,scale=0.34]{Z2j_over_gm2j_NLOMEPS_anti-kt-R5-Pt50_jetsHT500MHT150_HT.eps}
  \end{center}
  \caption{\baselineskip 14pt%
The $\HTj$ distribution. 
The left column shows
distributions for the Set 1 cuts, and the right column for the Set 2 cuts.
The plots are arranged and the curves are labeled as in \figs{fig:Jet1PT}{fig:Jet2PT}.
}
  \label{fig:HT}
\end{figure}

The situation changes slightly if we examine the $\HTj$ distributions,
shown in \fig{fig:HT}.
In both Sets 1 and~2, these are falling distributions
which peak at the cut values for $\HTj$
(300 and 500~GeV, respectively).  In Set 1, the NLO corrections range
from 10\% to 50\%, gradually increasing with increasing $\HTj$ through
550~GeV or so;
the shape of the distribution is thus modified at NLO.  
In contrast, the corrections
for Set 2 are fairly uniform across all $\HTj$, presumably because of the
larger $\HTj$ starting scale for the distribution.  Once again,
the corrections do not distinguish between $Z$ and $\gamma$ production,
so that the ratios for both Sets 1 and 2 have NLO corrections of about
10--15\%, with corrections relatively flat across all $\HTj$.
The NLO and ME+PS results for the ratio are again in good and uniform
agreement, to better than 10\%.
In both sets, the scale-dependence
bands are uniformly narrower at NLO than at LO.

\begin{figure}[t]
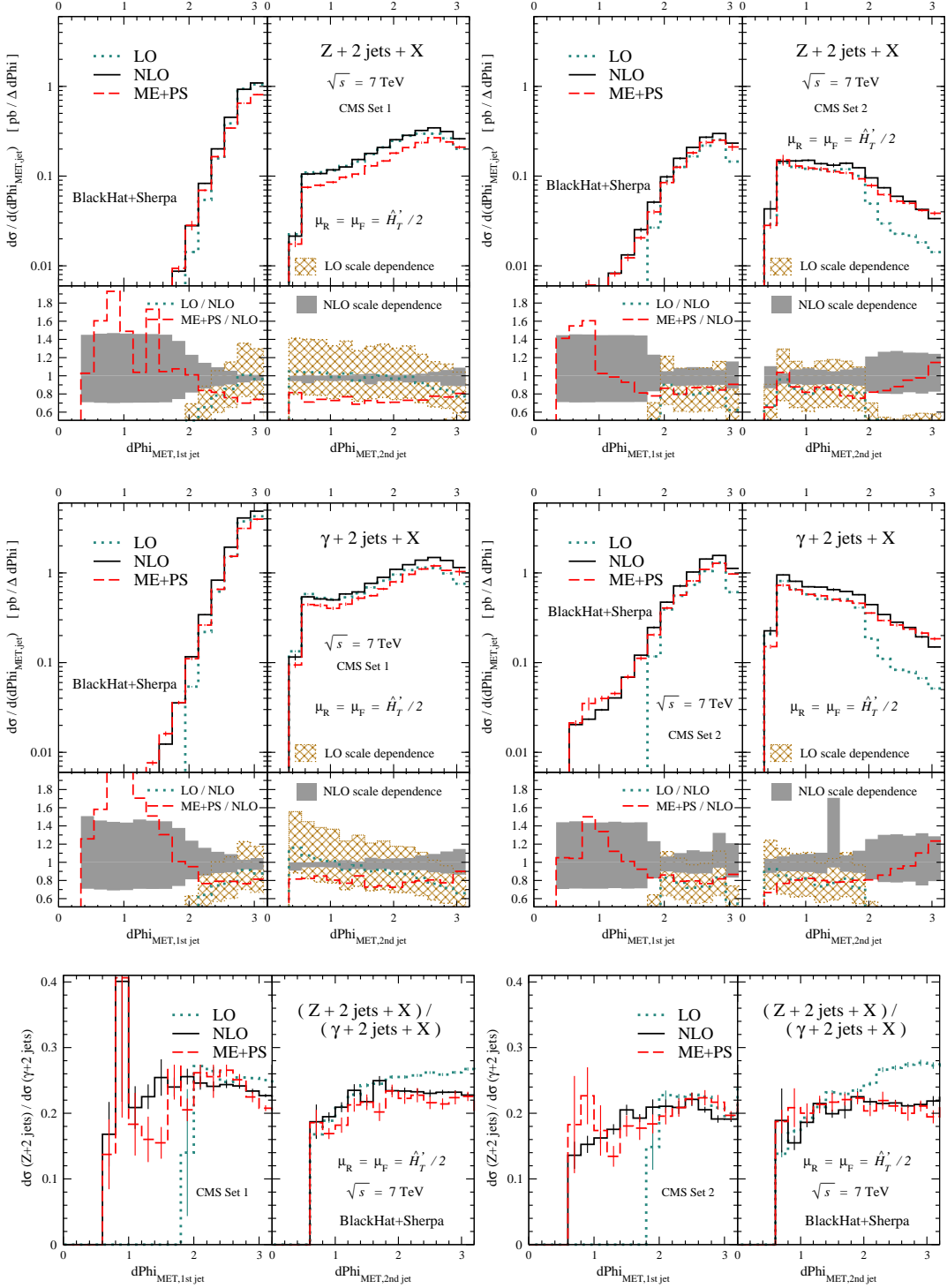

  \begin{center}
%
    \includegraphics[clip,scale=0.47]{Z2j-CMS-INC_anti-kt-R5-Pt50_jetsHT300MHT250_MHTlist_1_1_dphi2__An.eps}
    \includegraphics[clip,scale=0.47]{Z2j-CMS-INC_anti-kt-R5-Pt50_jetsHT500MHT150_MHTlist_1_1_dphi2__An.eps} \\
\vskip 10pt
    \includegraphics[clip,scale=0.47]{gm2j-CMS-INC_anti-kt-R5-Pt50_jetsHT300MHT250_MHTlist_1_1_dphi2__An.eps}
    \includegraphics[clip,scale=0.47]{gm2j-CMS-INC_anti-kt-R5-Pt50_jetsHT500MHT150_MHTlist_1_1_dphi2__An.eps} \\
\vskip 10pt
\hskip 5pt
    \includegraphics[clip,scale=0.47]{Z2j_over_gm2j_NLOMEPS_anti-kt-R5-Pt50_jetsHT300MHT250_MHTlist_1_1_dphi2__An.eps}
    \includegraphics[clip,scale=0.47]{Z2j_over_gm2j_NLOMEPS_anti-kt-R5-Pt50_jetsHT500MHT150_MHTlist_1_1_dphi2__An.eps}
  \end{center}
  \caption{\baselineskip 14pt%
The $\Delta\phi(\textrm{MET},\textrm{jet})$ distributions for the leading
two jets.  The left column shows the distributions for the Set 1 cuts,
and the right column for the Set 2 cuts.  Each column displays the
differential cross section for \Zjj-jet production (top), \gjj-jet
production (middle), and their ratio (bottom).  The left panel of each plot
shows the distribution for the leading jet, and the right panel for the second jet.
In the top and middle plots,
the upper panel shows the LO, NLO, and ME+PS results for the
distribution, and the lower panel shows the ratio to the central NLO
prediction, along with the LO and NLO scale-dependence bands.  }
  \label{fig:DPhi}
\end{figure}

\begin{figure}[t]
\begin{center}
  \subfigure[]{\label{fig:jetsMET@LO}%
    \includegraphics[clip,scale=0.75]{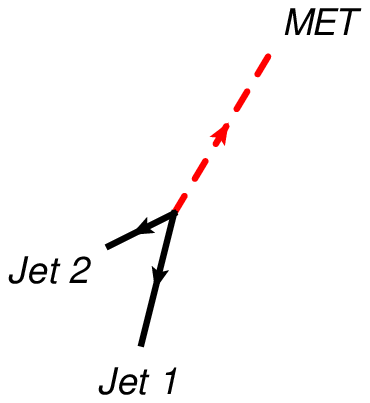}}\hskip 1in
  \subfigure[]{\label{fig:jetsMET@NLO}%
    \includegraphics[clip,scale=0.75]{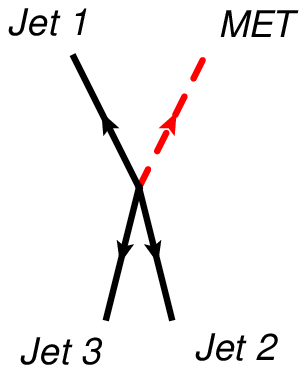}}
\end{center}
\caption{A typical configuration of jets and MET vector for the Set~1
cuts at (a) LO and (b) NLO.}
\label{fig:jetsMET}
\end{figure}

The distributions which show the most structure are those
in the azimuthal angle $\Delta\phi(\textrm{MET},\textrm{jet}_i)$
between the MET vector and the two jets, shown in \fig{fig:DPhi}.
The distribution for the first jet is peaked near $\Delta\phi = \pi$,
where the leading jet balances the MET vector while the second
jet has much smaller $p_T$.  
The distribution falls very steeply at smaller angle.
A typical LO configuration for the Set~1 cuts is shown in \fig{fig:jetsMET@LO}.
With only two partons and a missing energy vector from the $Z$ or $\gamma$,
transverse momentum conservation implies that the leading jet can never
get closer in angle than $\pi/2$ to the MET vector.
It can only approach this angle if the transverse momenta of both
jets are extremely large.  This configuration is heavily suppressed
by the parton distributions, to orders
of magnitude below what is shown in \fig{fig:DPhi}.
At NLO, this
kinematic constraint is relaxed, as it is possible in the real-emission
configurations for the second and third jets (the latter may
fall below the jet $p_T$ threshold)
to balance the leading jet and MET vector, as shown schematically
in \fig{fig:jetsMET@NLO}.  At this order,
the leading-jet distribution does indeed run all the way to the
$\Delta\phi(\textrm{MET},\textrm{jet})=0.5$ cut we impose.  The cross
section for angles less than $\pi/2$ is tiny, however, so although the
NLO corrections are very large in this region, they are of no practical
importance; this carries over to the large corrections seen in the
ratio of \Zjj-jet and \gjj-jet distributions.

The width of the peak region in the $\Delta\phi$
distribution for the leading jet
can be understood by considering a configuration in which the second jet
has $p_T\simeq 50$~GeV and $180$~GeV for Sets 1 and~2 respectively,
with the leading jet having the minimum $p_T$ needed to balance the
MET given a jet--jet $\Delta\phi=0.5$ (the interjet $\Delta R$ cut).  This
would lead to a peak around $\Delta\phi\simeq 3.05$ for Set 1, and $3.0$
for Set~2.  
However, the shapes of the ratios of
distributions do differ somewhat between the NLO and ME+PS
predictions for the peak region of the Set~1 distribution 
($\Delta\phi\sim 2.8$--$3.0$), so that the analysis would be more sensitive
to any systematic effects in the treatment of this region.

In the $\Delta\phi$ distribution for the second jet, we see 
different structure
in the small-angle region for \Zjj-jet and \gjj-jet production.
For Set 1, this region is subdominant in its contributions, although
not completely negligible, because the high MET cut favors configurations
in which both jet transverse momenta oppose the MET vector,
as shown in \fig{fig:jetsMET}(a).  In Set 2, however, the low MET cut allows
small values of $\Delta\phi(\textrm{MET},\textrm{jet}_2)$ to dominate.
In this region, the potential collinear
singularity between a photon and an outgoing quark forming the second
jet causes the cross section to start to rise at small angles.  This
rise is cut off by the $\Delta\phi(\textrm{MET},\textrm{jet})$ cut,
but does distinguish $Z$ production from photon production.  For
both Sets 1 and~2, the shapes of the ratios of
distributions are similar at LO, at NLO, and in the ME+PS results,
so that we can be confident the $Z$-to-$\gamma$ difference is captured
properly, and that the analysis should not be overly sensitive to
the precise value chosen for the $\Delta\phi(\textrm{MET},\textrm{jet})$ cut.
The shapes of the NLO and ME+PS distributions (though not the LO one)
are quite similar as well in the peak region, so that the analysis
should be robust under systematic effects treating different angular
regions in $\Delta\phi(\textrm{MET},\textrm{jet}_2)$ somewhat differently.

To assess the overall uncertainty, we see that the ME+PS prediction
for the total cross section with the Set 1 cuts is approximately 7\%
lower than the NLO prediction, while for the Set~2 cuts it is
approximately 5\% higher.  In addition, we should allow for a
difference of up to 1\% in the prediction for the fixed-cone isolation
compared to the Frixione-style isolation, as discussed in
section~\ref{InclusivePhotonSection}. This suggests that the
NLO predictions for the ratios used by the CMS collaboration should be
accurate to within 10\%.

We have not considered the effect of possible 
electroweak Sudakov logarithms.
  Based on refs.~\cite{ElectroweakLogs}, we
estimate that for Set 1, the effect will be under 5\% and
smaller for Sets 2 and 3.  Our study considers production of
an (exclusive) 
electroweak boson accompanied by at least two jets rather than one;
 we expect the electroweak
effects to be of similar magnitude for similar boson transverse momenta.  
If one were to raise the MET cut, these
electroweak corrections increase, and would need to be taken into account
to ensure the reliability of the prediction
for the \Zjj-jet to \gjj-jet ratio.  Were the MET cut to be raised to
500~GeV, for example, the virtual electroweak corrections would grow to
about 10\%.  The effects of radiating a second electroweak gauge boson
into the final state~\cite{BaurEW} should also be considered.
These questions merit further investigation.


\section{Conclusions and Outlook}

Data-driven estimates of backgrounds to new physics signals offer a
powerful means for avoiding a reliance on Monte Carlo estimates of
backgrounds.  Even in such data-driven methods, theoretical input is
usually required to provide conversion factors from one process to
another, and estimates of uncertainties in these factors, as well as
of extrapolations from control to signal regions.

In this paper, we studied the theoretical aspects of using \gp{}jets
data to estimate the missing-$E_T$\,+\,jets background to new-physics
searches.  We focused on the comparison of boson production in
association with two or more jets, computing the relevant differential
cross sections and ratios in NLO QCD.  Because the scale uncertainty
is quite small in these ratios, we used the difference between ME+PS
results and the NLO ones as an estimate of theoretical uncertainty.
(We did not study the uncertainty due to parton distribution
functions, but it is expected to be small as well.) Our study used the
Frixione isolation criterion to compute the prompt-photon cross
sections; a comparison with the fixed-cone isolation for isolated
prompt-photon production indicates that the resulting shift should be
less than 1\% in the high-$p_T^\gamma$ region of interest.
We did not include the effects of electroweak Sudakov logarithms,
and while these will become significant for higher MET cuts, we expect
the effects to remain small for the cuts used in this study.

We find that the conversion between photons and $Z$ bosons has less
than a 10\% theoretical uncertainty for events with two associated
jets.  In the future, it should be feasible to extend this study to
three associated jets.  The small uncertainty we find should make it
possible for the photon channel to provide a competitive determination
of the Standard-Model missing-$E_T$\,+\,jets background for many years
to come.

\vskip .3 cm 

\centerline{\bf Note added}

Since the appearance of our paper, CMS presented a new physics search
based on an event signature of at least three jets accompanied
by large missing transverse momentum~\cite{CMSPhotons}.  No excess
events were observed above the background. The irreducible background
from $Z$ bosons decaying into a $\nu \bar \nu$ pair was estimated by
converting a measurement of photons accompanied by jets into a
prediction of this background.  Our paper provided the estimate
of the theoretical uncertainty for this conversion.

\section*{Acknowledgements}

We thank Joe Incandela, Sue-Ann Koay, Steven Lowette, Roberto Rossin,
Piet Verwilligen and Mariarosaria D'Alfonso, for suggesting this
analysis, for bringing CMS's study to our attention, and for extensive
discussions.  We thank Andr\'e David and David d'Enterria for
providing the \jetphox{} predictions in
ref.~\cite{CMSInclusivePhoton}.  We thank Werner Vogelsang for
providing a copy of his NLO isolated prompt-photon code, and
Jean-Philippe Guillet for providing a copy of the \jetphox{} code, and
for discussions pertaining to it. We also thank Ed Berger, Thomas
Gehrmann, Bernd Kniehl, Bill Marciano, Michael Peskin, Joe Polchinski,
Alberto Sirlin and Bennie Ward for helpful discussions.  We especially
thank Carola Berger and Tanju Gleisberg for their contributions during
early stages of this work.  This manuscript was completed during a
visit to the Kavli Institute for Theoretical Physics, and we thank it
for its hospitality.  This research was supported by the National
Science Foundation under Grant No. NSF PHY05--51164, and by the US
Department of Energy under contracts DE--FG03--91ER40662,
DE--AC02--76SF00515 and DE--FC02--94ER40818.  DAK's research is
supported by the European Research Council under Advanced Investigator
Grant ERC--AdG--228301.  This research used resources of Academic
Technology Services at UCLA and of the National Energy Research
Scientific Computing Center, which is supported by the Office of
Science of the U.S. Department of Energy under Contract
No. DE--AC02--05CH11231.



\begin{thebibliography}{99}

\bibitem{CMS-photon-note}
CMS Collaboration, \textit{``Data-Driven Estimation of the Invisible Z 
Background to the SUSY MET Plus Jets Search''}, CMS PAS SUS-08-002 (2008),
unpublished.

\bibitem{CMSMET}
CMS Collaboration, \textit{``Search for new Physics at CMS with Jets and 
Missing Momentum''}, CMS PAS SUS-10-005 (2010), unpublished.

\bibitem{W3j}
C.~F.~Berger, Z.~Bern, L.~J.~Dixon, F.~Febres Cordero, D.~Forde,
T.~Gleisberg, H.~Ita, D.~A.~Kosower and D.~Ma\^{\i}tre,
Phys.\ Rev.\  D {\bf 80}, 074036 (2009)
[0907.1984 [hep-ph]].

\bibitem{W4j}
C.~F.~Berger, Z.~Bern, L.~J.~Dixon, F.~Febres~Cordero, D.~Forde, 
T.~Gleisberg, H.~Ita, D.~A.~Kosower and D.~Ma\^{\i}tre, 
Phys.\ Rev.\ Lett.\  {\bf 106}, 092001 (2011)
[1009.2338 [hep-ph]].

\bibitem{BlackHatZ3jet}
C.~F.~Berger, Z.~Bern, L.~J.~Dixon, F.~Febres~Cordero, D.~Forde,
T.~Gleisberg, H.~Ita, D.~A.~Kosower and D.~Ma\^{\i}tre,
Phys.\ Rev.\ D {\bf 82}, 074002 (2010)
[1004.1659 [hep-ph]].

\bibitem{Wpolarization}
Z.~Bern, G.~Diana, L.~J.~Dixon, F.~Febres Cordero, D.~Forde, T.~Gleisberg, 
S.~H{\"o}che, H.~Ita, D.~A.~Kosower, D.~Ma\^{\i}tre and K.~Ozeren, 
1103.5445 [hep-ph].

\bibitem{BlackHatI}
C.~F.~Berger, Z.~Bern, L.~J.~Dixon, F.~Febres~Cordero, D.~Forde, H.~Ita,
D.~A.~Kosower and D.~Ma\^{\i}tre,
Phys.\ Rev.\ D {\bf 78}, 036003 (2008)
[0803.4180 [hep-ph]].

\bibitem{BlackHatII}
C.~F.~Berger, Z.~Bern, L.~J.~Dixon, F.~Febres Cordero, D.~Forde, T.~Gleisberg,
H.~Ita, D.~A.~Kosower and D.~Ma\^{\i}tre,
Phys.\ Rev.\ Lett.\  {\bf 102}, 222001 (2009)
[0902.2760 [hep-ph]].

\bibitem{Amegic}
F.~Krauss, R.~Kuhn and G.~Soff,
JHEP {\bf 0202}, 044 (2002)
[hep-ph/0109036];\\
%
T.~Gleisberg and F.~Krauss,
Eur.\ Phys.\ J.\  C {\bf 53}, 501 (2008)
[0709.2881 [hep-ph]].

\bibitem{Sherpa}
T.~Gleisberg, S.~H{\"o}che, F.~Krauss, A.~Schalicke, S.~Schumann and 
J.~C.~Winter,
JHEP {\bf 0402}, 056 (2004)
[hep-ph/0311263];\\
T.~Gleisberg, S.~H\"{o}che, F.~Krauss, M.~Sch\"{o}nherr, S.~Schumann,
F.~Siegert and J.~Winter,
JHEP {\bf 0902}, 007 (2009)
[0811.4622 [hep-ph]].

\bibitem{HoechePhoton}
S.~H{\"o}che, S.~Schumann and F.~Siegert,
Phys.\ Rev.\  D {\bf 81}, 034026 (2010)
[0912.3501 [hep-ph]].

\bibitem{Frixione}
S.~Frixione,
Phys.\ Lett.\  B {\bf 429}, 369 (1998)
[hep-ph/9801442].

\bibitem{PhotonFragmentation}
L.~Bourhis, M.~Fontannaz and J.~P.~Guillet,
Eur.\ Phys.\ J.\  C {\bf 2}, 529 (1998)
[hep-ph/9704447].

\bibitem{CMSInclusivePhoton}
V.~Khachatryan {\it et al.}  [CMS Collaboration],
Phys.\ Rev.\ Lett.\  {\bf 106}, 082001 (2011)
[1012.0799 [hep-ex]].

\bibitem{CS}
S.~Catani and M.~H.~Seymour,
Nucl.\ Phys.\  B {\bf 485}, 291 (1997)
[Erratum-ibid.\  B {\bf 510}, 503 (1998)]
[hep-ph/9605323].

\bibitem{Zqqgg}
Z.~Bern, L.~J.~Dixon and D.~A.~Kosower,
Nucl.\ Phys.\  B {\bf 513}, 3 (1998)
[hep-ph/9708239].

\bibitem{Zqqqq}
Z.~Bern, L.~J.~Dixon, D.~A.~Kosower and S.~Weinzierl,
Nucl.\ Phys.\  B {\bf 489}, 3 (1997)
[hep-ph/9610370].

\bibitem{OtherZpppp}
J.~M.~Campbell, E.~W.~N.~Glover and D.~J.~Miller,
Phys.\ Lett.\  B {\bf 409}, 503 (1997)
[hep-ph/9706297].

\bibitem{qqggg}
Z.~Bern, L.~J.~Dixon and D.~A.~Kosower,
Nucl.\ Phys.\  B {\bf 437}, 259 (1995)
[hep-ph/9409393].

\bibitem{Kunsztqqqqg}
Z.~Kunszt, A.~Signer and Z.~Tr\'ocs\'anyi,
Phys.\ Lett.\  B {\bf 336}, 529-536 (1994)
[hep-ph/9405386].

\bibitem{SignerPhoton}
A.~Signer,
Phys.\ Lett.\  B {\bf 357}, 204 (1995)
[hep-ph/9507442].

\bibitem{MCFM}
J.~M.~Campbell and R.~K.~Ellis,
Phys.\ Rev.\  D {\bf 65}, 113007 (2002)
[hep-ph/0202176].

\bibitem{Comix}
T.~Gleisberg and S.~H{\"o}che,
JHEP {\bf 0812}, 039 (2008)
[0808.3674 [hep-ph]].

\bibitem{CSShower}
S.~Schumann and F.~Krauss,
JHEP {\bf 0803}, 038 (2008)
[0709.1027 [hep-ph]].

\bibitem{PhotonShower}
S.~H{\"o}che, S.~Schumann and F.~Siegert,
Phys.\ Rev.\  D {\bf 81}, 034026 (2010)
[0912.3501 [hep-ph]].

\bibitem{KniehlLonnblad}
B.~A.~Kniehl and L.~L\"onnblad,
DESY 92--032 (1992),
in {\it Proceedings of the Annecy Photon Workshop (1991)}.

\bibitem{MarcianoAlpha}
A.~Czarnecki and W.~J.~Marciano,
Phys.\ Rev.\ Lett.\  {\bf 81}, 277 (1998)
[hep-ph/9804252].

\bibitem{antikT}
M.~Cacciari, G.~P.~Salam and G.~Soyez,
JHEP {\bf 0804}, 063 (2008)
[0802.1189 [hep-ph]].

\bibitem{LesHouchesDiscretizedFrixioneStudy}
J.~R.~Andersen {\it et al.}  [SM and NLO Multileg Working Group],
1003.1241 [hep-ph] (page 94).

\bibitem{JetPhoX}
S.~Catani, M.~Fontannaz, J.~Ph.~Guillet and E.~Pilon,
JHEP {\bf 0205}, 028 (2002)
[hep-ph/0204023]; \\
%
P.~Aurenche, M.~Fontannaz, J.~Ph.~Guillet, E.~Pilon and M.~Werlen,
Phys.\ Rev.\  D {\bf 73}, 094007 (2006)
[hep-ph/0602133].

\bibitem{Vogelsang}
L.~E.~Gordon and W.~Vogelsang,
Phys.\ Rev.\  D {\bf 50}, 1901 (1994).

\bibitem{CT10}
H.-L.~Lai, M.~Guzzi, J.~Huston, Z.~Li, P.~M.~Nadolsky, 
J.~Pumplin and C.-P.~Yuan,
Phys.\ Rev.\  D {\bf 82}, 074024 (2010)
[1007.2241 [hep-ph]].

\bibitem{FrixioneVogelsangPolarized}
S.~Frixione and W.~Vogelsang,
Nucl.\ Phys.\  B {\bf 568}, 60 (2000)
[hep-ph/9908387].

\bibitem{ElectroweakLogs}
E.~Maina, S.~Moretti and D.~A.~Ross,
Phys.\ Lett.\  B {\bf 593}, 143 (2004)
[Erratum-ibid.\  B {\bf 614}, 216 (2005)], 
[hep-ph/0403050];\\
%
J.~H.~Kuhn, A.~Kulesza, S.~Pozzorini and M.~Schulze,
JHEP {\bf 0603}, 059 (2006)
[hep-ph/0508253].

\bibitem{BaurEW}
U.~Baur,
Phys.\ Rev.\  D {\bf 75}, 013005 (2007)
[hep-ph/0611241].

\bibitem{CMSPhotons}
S.~Chatrchyan {\it et al.}  [CMS Collaboration],
1106.4503 [hep-ex].

\end{thebibliography}
\end{document}